\documentclass[%
 prx,
rsi,%
 amsmath,amssymb,
 reprint,%
superscriptaddress
]{revtex4-1}

\usepackage{graphicx}
\usepackage{dcolumn}
\usepackage{bm}
\usepackage{color}

\begin{document}

\title{Model for growth and morphology of fungal mycelium}

\author{Bhagyashri Shinde}
\author{Shagufta Khan}
\author{Sudipto Muhuri}
\email{sudipto@physics.unipune.ac.in}
\affiliation{Department of Physics, Savitribai Phule Pune University, Pune, India}

\date{\today}

\begin{abstract}

 We present a minimal driven lattice gas model which generates the morphological characteristics associated with single colony mycelium arising from the growth and branching process of fungal hyphae, which is fed by a single source of nutrients.
We first analyze the growth and transport process in the primary hypha modeled as a growing 1-d lattice, which is subject to particle (vesicle) loss due to presence of dynamically created branching sites. We show that the spatial profile of vesicles along the growing lattice is an exponential distribution, while the length grows logarithmically with time.  We also find that the probability distribution of length of the hypha tends to a Gaussian distribution function at late times. In contrast, the probability distribution function of the time required for growth to a specific length tends to a broad  log-normal distribution.  
We simulate the resultant 2-d morphology generated by the growing primary hypha, quantifying the motility behavior and morphological characteristics of the colony. Analysis of the temporal behavior and morphological characteristics of the resultant 2-d morphology reveals a wide variability of these characteristics which depend on the input parameters which characterize the branching and elongation dynamics of the hyphae. By calibrating the input parameters for our model, we make some quantitative comparison of the predictions of our model with the observed experimental growth characteristics of fungal hyphae and the morphological characteristics of single colony fungal mycelium. 
\end{abstract}

\keywords{Fungi}
\maketitle

\section{Introduction}
Fungi is an integral part of the nutrient cycle in ecosystems  and they serve as  important model systems for genetic research due to their relative simplicity as an eukaryotic cellular system \cite{deacon, genome}.  

Fungi exhibit filamentous growth process. The basic filament structure of Fungi- the {\it hyphae}, grow by means of extension and branching at the their tip, giving rise to a multicellular complex network -{\it mycelium} \cite{deacon,cell}. In order to develop a comprehensive understanding of the morphology of the entire {\it mycelium}, it is pertinent to develop an understanding of the transport and branching process of the individual fungal hypha and the morphological features of colony formed as a result of such a growth process. Depending on the scale of description of the growth phenomenon, theoretical models for growth and transport in individual hypha and fungal colonies belong to the categories of tip-scale models \cite{vsc1,vsc2}, intermediate scale models \cite{plos,intermediate,sugden-pre,trinci} and macroscale models \cite{boswell, genome}. The tip-scale models such as vesicle supply center (VSC) model focus on aspects of extension  and shape of the hyphal tip \cite{vsc1,vsc2} and the typical length scales of description is $\sim 100\mu m$.  For this model the connection of the growth or branching process with the process of vesicle supply from the sub-apical region of the fungi is not explicitly taken into account \cite{sugden-pre, plos}. On the other hand the macroscale models focus on description in terms of effective interaction of fungal colony as a whole with the environment, without explicitly taking into account the metabolic processes at play for the individual hypha that constitute the colony \cite{mycelium, boswell}. The model that we study belongs to the category of intermediate scale description, wherein, we focus on motor driven process of growth and branching in individual hyphae which is fed by a single source of nutrients and generates  a {\it single colony mycelium}. Much of  the previous work in the domain of such intermediate scale description of fungal growth has taken recourse to continuum models and focused on the details of bio-mechanics, without explicitly considering the role of molecular motors \cite{plos,boswell}. In contrast we investigate the role of molecular motors using a model of driven lattice gas. 

Driven lattice gas models have provided useful description for plethora of  biological processes encompassing transport across biomembranes~\cite{chou}, dynamics of ribosomes in m-RNA~\cite{mcdonald}, motor driven intracellular transport~\cite{freypre, freylet, menon, ignapre,epl-abhishek}, transport in fungal hyphae ~\cite{sugden-pre,sugden-jstat, sm-epl} and other driven phenomenon \cite{santen, sm-epl, evans,schutz,popkov,evans1,evans2, lakatos,shaw,derrida}. In particular, a driven lattice gas model - Dynamically extending exclusion process (DEEP)  has been  adopted for modeling extension of fungal hyphae \cite{sugden-pre,sugden-jstat}. For DEEP, particles hop unidirectionally on the lattice, interacting with other particles via hard-core repulsion, and the particles at the growing end of the lattice dynamically create a new lattice site, leading to the overall extension of the lattice \cite{sugden-jstat,sugden-pre}. While this model has been able to relate the growth of the fungal hyphae with the supply of nutrients, one inadequacy has been that the process of branching and related loss of the transported particles has not been taken into account. For instance for aerial hyphae of {\it sporangiophore}, branching is indeed observed. Further, the elongation process of the individual hypha is such that there is a slow down of their growth rates as their length increases \cite{plos,gruen}. This is in contrast to the results obtained for DEEP, which predicts a growth rate of the hypha that is uniform, without any slowdown \cite{sugden-pre,sugden-jstat}. 

Our minimal model is a generalization of the Dynamically extending exclusion process (DEEP), encompassing the process of both linear growth and branching. Also in contrast to continuum bio-mechanical models \cite{plos}, which involves many parameters, our coarse grained description involves  very few input parameters e.g; nutrient supply rate, growth rate at the tip, branching rate, and flow rate of nutrients to the  different branches of the growing lattice. One experimental system that serves  as a paradigmatic example is the  fungus {\it Nuerospora Crassa} \cite{gero,neurospora}. For this system, the material necessary for growth of the hyphae is packaged as vesicles and supplied by a single source located in the sub-apical region of the fungi \cite{gero,neurospora}. The transport of these vesicles is done by molecular motors, which walk along the parallel array of microtubule filaments, carrying these vesicles to the tip of the hypha \cite{mt1,mt2}. In the apical region of the tip, the intracellular organelle- {\it Spitzenkorper} is involved in synthesis of new cell wall from the vesicles, leading to growth \cite{gero}. Apart from linear elongation process of the hyphae, lateral branching process  of the fungal hyphae is also observed \cite{neurospora,gruen,plos}. The process of branching and elongation of a single primary hypha, fed by a single source of nutrient generates a single colony mycelium. We study the morphological characteristics and growth pattern of such colonies on a surface using a 2-d minimal 
model. 2-d growth processes have been studied  for randomly branched polymers \cite{polymer,rosa1,rosa2} and growth of cell colonies \cite{eden}. In this context, variants of {\it lattice animal} models such as Eden Aggregation process have been used to characterize the growth processes on discrete lattices \cite{eden, eden-prl, eden1,eden2, dhar1,dhar2, latticeanimal1}, while continuum models such as Flory Theory for branched polymers have been developed to study polymerization process under different conditions \cite{polymer,rosa1,rosa2}.  The connections between these discrete lattice animal models with the polymerization models have also been studied \cite{polymer,rosa1,rosa2, lubensky1, lubensky2}. There is however one crucial distinction of these models  with the model that we study in this article. In particular for our model, the elongation and branching process at individual sites of the colony is coupled to supply of nutrients at those sites, and is in contrast to growth process of branched polymers and lattice animals for which growth process at the sites are not subject to such a constraint. 

In Section~\ref{sec:model} we first describe the 1-d minimal model for the growth and transport in the primary hypha, specifying the dynamical rules for growth and set up the corresponding equations of motion for the system. Further we also specify the dynamical rules governing the development of the 2-d mycelium arising from the branching process of the hyphae.

In Section~\ref{sec-1d} we obtain the mean field (MF) analytical solutions of the length of the individual primary hypha as a function of time as well as the spatial profile of the transported cargo along the hypha. We then compare these results with the Monte Carlo (MC) simulation results. By performing MC simulations, we also obtain the probability distribution of lengths of the primary hypha, and the probability distribution of the time required for growth up to a specified length of the hypha. 

In Section~\ref{sec-2d} we discuss the results for the spatial and temporal features associated with single colony mycelium that is generated from the elongation and branching process of the primary hyphae. 

In Section~\ref{sec-expt} we calibrate the input parameters of the model to reproduce the quantitative measures associated with the growth characteristics of the individual fungal hypha and single colony mycelium, that are observed in experiments.

In Section~\ref{sec-conclusion} we summarize our results and discuss the insights gained from the findings of this model in understanding the growth and morphological characteristics of single colony mycelium.  

\section{Model}
\label{sec:model}

\begin{figure}[h]
\centering
 \includegraphics[width=0.9\linewidth]{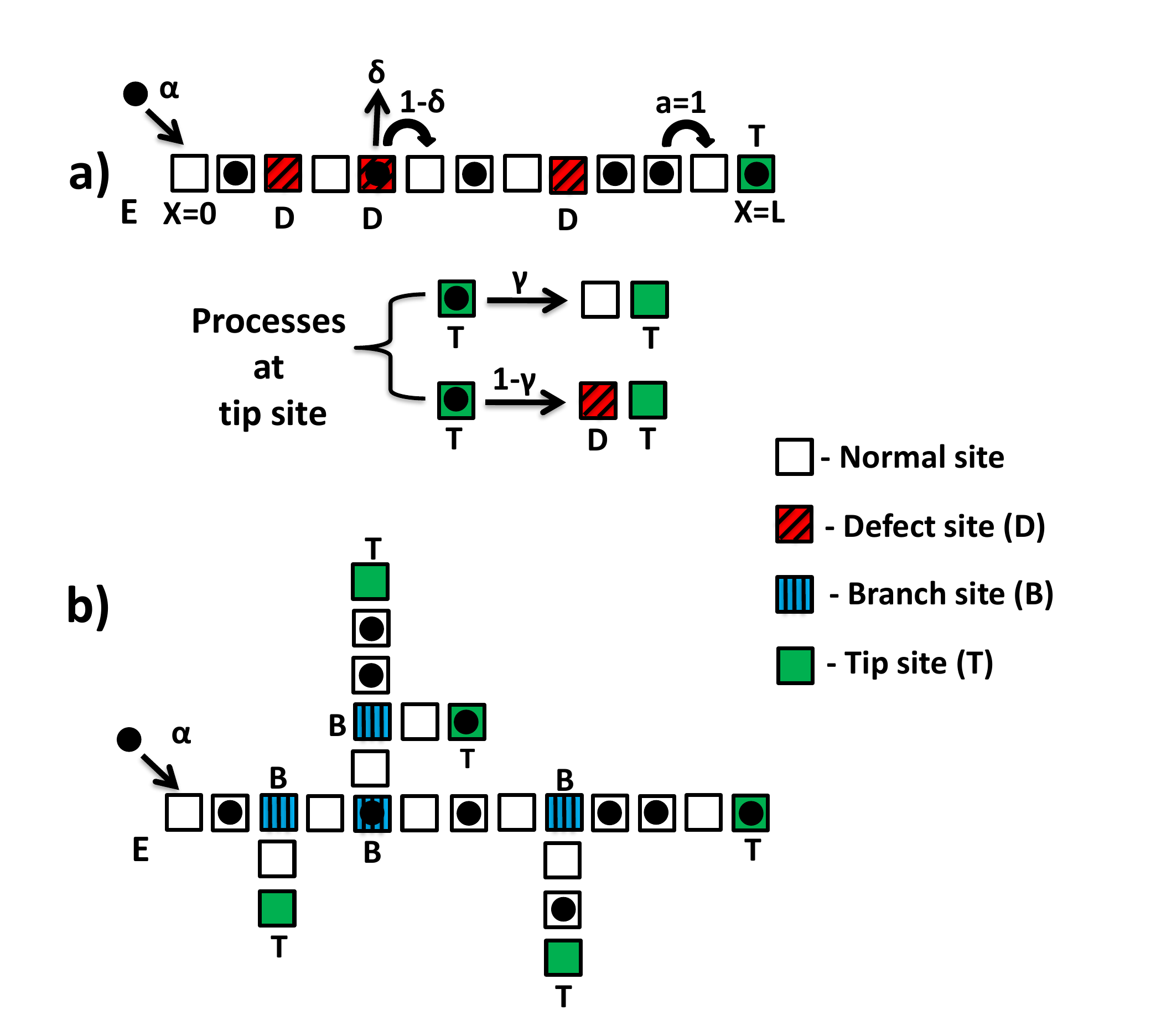}
\begin{center}
\caption{Schematic  of the dynamical processes for (a) Growing primary hypha is represented as 1-d growing lattice. With a rate $\gamma$, an occupied tip site (T) becomes two normal empty site while with rate $1 -\gamma$ two empty lattice sites are created with the preceding site transforming into a defect site (D). The particle entry rate at first site (E) is $\alpha$, particle hopping rate is $1$ for a normal (non-defect) site and loss rate of particles at defect site is  $\delta $.  (b) Single colony mycelium represented as a 2-d lattice with entry rate $\alpha$, branch site creation rate $1-\gamma$, hopping rate of $1$ for a normal (non-branching) site and particle flow rate to secondary branch is $ \delta$. For the branching sites, hopping rate on the same branch is $1- \delta$ and overall growth rate at tip is $1$. Here (E) denotes the site of entry, (T) denotes tip site and (B) a branching site.}
\end{center}
 \label{fig1}
\end{figure}

In general, process of  linear growth and lateral branching (in same plane)  of the {\it primary} fungal hypha would generate a  2-d morphology. We first describe the 1-d model for growth and transport behavior of the primary hypha. Subsequently we would discuss the generalized model for describing the 2-d morphology of the hyphal colony emerging due to the lateral branching process and linear extension of the fungal hyphae. 

\subsection { Effective 1-d model for growth for primary hypha} 
\label{1-d}

 We consider the initial state of the fungal hypha to be an object of 1-d spatial extension with an initial specified length $\epsilon$ and which we refer to as the primary hypha. From the perspective of the linear extension of this {\it primary} hypha, the effect of branching manifests as a process of material loss along the hyphae during the process of its growth. We represent this growing primary fungal hypha as a 1-d discrete lattice. At any instant of time $t$, the total number of lattice sites is $N$ corresponding to a length $L = \epsilon N$ (Fig.1). Vesicles required for growth are represented by particles, which are transported by molecular motors along the lattice from left to right. First we describe the dynamics of the particles at the lattice boundaries. At left end of the lattice, at site $i =1$, particles enter the lattice with a rate $\alpha$ provided the boundary site is vacant. At the boundary site on the right, at site $i = N$, which corresponds to the tip of the hypha, two processes can occur: (a) If the tip site at $i =N$ is occupied by a particle then with a rate $\gamma$ an additional {\it empty}  lattice site to the right is created, and the site  $i = N$ also become {\it empty}. This process leads to overall increment of the length of the lattice from from $L$ to  $L + \epsilon$, while the total number of lattice sites is incremented from $N$ to $N +1$. This dynamics of lattice growth is exactly same as that of DEEP \cite{sugden-pre,sugden-jstat}. (b) With a rate $ 1 - \gamma$, an {\it occupied}  tip site $i =N$ not only creates a new empty lattice site to the right, but the lattice at $i=N$ becomes an empty  {\it defect site}.  The overall extension rate of the lattice is $1$. These defect sites correspond to the branching sites that would get created dynamically. If one considers the full 2-d morphology resulting from this growth process then the defect sites would correspond to the branching sites for which the vesicle supply from the static end of hyphae would be split between the two branches, leading to eventual growth of these two separate branches. However from the perspective of the primary hypha, the effect of branching would manifest as loss of vesicle at the {\it defect} sites. Thus effectively at any instant of the time, the 1-d lattice, representing the primary hypha would comprise of bulk normal bulk sites (N), {\it defect sites} (D), apart from the boundary sites at the growing tip (T) and boundary site at the non-growing end (E), which corresponds to the site where particles enter the lattice. Particles in the bulk of the lattice interact with each other via excluded volume effect, thus restricting the maximum occupancy per lattice site to 1. For any normal bulk site $i$, the particles hop unidirectionally to the adjacent site to the right, with rate $1$ provided the site at $i+1$ is vacant. At the defect site, with a rate $\delta$ there is a loss of particle from the site, while with a rate $1-\delta$, the particles hop to the site $i+1$, provided it is empty. This completes the specification of the dynamics of the 1-d model. All the dynamic processes for this dynamic lattice model are schematically depicted in Fig.~1(a). 

\subsection { 2-d model for single colony mycelium} 
\label{2-d}

We consider the growth process of single colony on a square lattice, arising due to the branching and elongation process of the fungal hyphae constituting the colony. We consider an initial starting configuration of the colony which is a fungal hypha, and represented by a 1-d lattice of length $\epsilon$ placed along the $x$ axis at the origin. The processes of particle entry at the left boundary is same as that of the 1-d model, with the particles entering at left end of the lattice located at $(0,0)$, with a rate $\alpha$ provided the site is vacant. If there is particle at the tip site, two processes occur: (i) {\it Pure Linear Extension}:  With a rate $\gamma$, a new lattice site is created in the original direction of movement of the particle, provided that the site is not occupied by the colony itself. The underlying assumption made here is that presence of the colony obstructs and arrests the linear growth process of an hyphal tip. (ii) {\it Branching and extension}: With a rate $ 1- \gamma$, two new empty lattice sites are created- an empty lattice in the original direction of movement of particle, and an empty site  perpendicular to it. One of two perpendicular direction is chosen randomly with equal probability. Again this process is subjected to the constraint that it occurs {\it only if} the adjacent site in the direction of growth and the adjacent sites in the perpendicular direction are not occupied by the colony itself. Such branching process would transform the tip site (T) at time $t$ into a {\it branching site} (B)  at time $t+1$. It would also create two new tip sites in the direction of original movement of particle and in a direction perpendicular to it. Both these processes would lead to growth of the single colony comprising of these hyphae. Thus at any instance of time any site in the bulk of the colony would either be a {\it branching} site (B) or a {\it normal} site N ( a site without any branching). At any instance a particle on a normal site will hop unidirectionally with rate $1$ to the adjacent site in the original direction of motion of particle provided it is empty. For a particle on a  branching site, with a rate $\delta$ the particle would hop on to the other branch site provided the adjacent site on the branch is empty, while with rate $1 - \delta$ it would continue to hop to adjacent site in the original direction of movement provided that site is empty.  A typical configuration is schematically depicted in Fig.1(b).

\subsection{MC simulation of the process}

 For determining the density and current profiles on hypha, Monte Carlo (MC) simulations have been performed to simulate the various processes for the 1-d model described in subsections~\ref{1-d} using the procedure of random-sequential update of the sites \cite{parallel1, parallel2}. In this procedure, at any given time step, a site is chosen at random with equal probability and the lattice site is updated according to the rules of dynamics specified in subsection~\ref{1-d} .  Each Monte Carlo unit of time $\Delta t_{mc}$  corresponds to an interval of real time  $\Delta t$, such that  $\Delta t = \Delta t_{mc} / N(t)$. We note that since the lattice length is growing, therefore, $N$ changes with time.  For simulating the configurations for the 1-d model, the initial starting configuration is a 1-d lattice with number of sites (N = 1) which is occupied by a particle at $x =0$. For the 2-d model, the initial starting configuration is also a 1-d lattice with a particle occupying site specified by coordinates $x=0, y =0 $, and the initial growth direction being along $+x$ direction. We adopt the synchronous update procedure ( also refered to as fully parallel update procedure) for updating the lattice sites for the 2-d model \cite{tilstra, parallel1, parallel2}. For this update procedure, at any instance of time $t$, for a given configuration, {\it all the lattice sites are updated simultaneously} according to the rules of dynamics specified in subsection~\ref{2-d} \cite{parallel2}. One distinct advantage of performing MC simulations using Synchronous update procedure, is that it is  relatively faster than random sequential update procedure \cite{parallel2} and has been used in context of modeling traffic flow \cite{parallel1}.  Ensemble averaging is done typically for 10000 samples for the 1-d model and for 100 samples for the 2-d model, starting with the same initial configuration. For the 1-d model, we have additionally also simulated the growing lattice  using a fully parallel update proedure to characterize the temporal behaviour of the growing lattice and compared the simulation results with the one obtained using a random-sequential update procedure.

\section{Characteristics of single hypha}
\label{sec-1d}

Having described the 1-d model for growth of the primary hypha in section \ref{1-d}, we now discuss the growth characteristics that we obtain for this minimal model.
\subsection{Survival probability and density profile}
For the 1-d model for hyphal growth, we consider an ensemble of similarly prepared system. We define survival probability $P( i | j )$ (with $i > j$) as the probability of a particle in site j to reach site i without leaving the 1-d lattice.
\begin{equation}
P(i | j ) = \displaystyle\prod_{s=j}^{i-1} P(s + 1 | s) P(j)
\label{pij}
\end{equation}  
where, $P(s + 1 | s)$ is the conditional probability for a particle at site $s$ to reach site $s+1$, and $P(j)$ is the probability of occupancy of site $j$.

On averaging over the ensemble, the average survival probability for a particle at site $s$ to reach a site $s+1$,  $\langle P(s + 1 | s) \rangle$ can be expressed in terms of the probability of occurrence of defect/normal sites and their respective survival probabilities as follows:
\begin{equation}
\langle P( s+1 | s) \rangle = P_{N}(s + 1 |s)P(N) +  P_{D}(s + 1 |s)P(D)
\end{equation}
where, $P(D)$ is the probability that the site $s$ is a {\it defect} site, $P(N)$ is the probability that it is a {\it normal} site, $P_{D}(s + 1 |s)$ is the survival probability of the particle at site $s$ to reach site $s+1$
 if the site $s$ is defect site, and $P_{N}(s + 1 |s)$ is the survival probability of the particle at site $s$ to reach site $s+1$ if the site $s$ is normal site. 
It therefore follows that, $\langle P( s+1 | s) \rangle = \gamma + (1- \delta)(1-\gamma)$, and the expression for average survival probability   $\langle P(i | j) \rangle$ assumes the form, 

\begin{equation}
\langle P(i | j) \rangle  = P(j){ \left[ 1- \delta +\delta \gamma \right]}^{(i - j)} 
\label{pij}
\end{equation}  
The distance of separation between the site $i$ and site $j$ in terms of lattice spacing $\epsilon$ can be expressed as  $x = (i - j) \epsilon$. Then the expression for the average probability of a particle surviving a distance $x$ is,

 \begin{equation}
P(x) = \exp \left( \frac{K x} {\epsilon} \right)
\label{px}
\end{equation}  

where $ K= ln [ 1- \delta + \delta \gamma ]$. 
The  steady state average occupancy, $\rho(0)$  at the left end of the lattice at $ x= 0$ is $\alpha$. 

When the excluded volume effect is ignored for the particle hopping process  in the growing lattice, the steady state average occupancy $\rho(x)$  maybe expressed as $\rho(x) = \rho(0) P(x)$ and it follows from Eq. (\ref{px}) that the approximate expression for the average occupancy $\rho(x)$ is, 
\begin{equation}
\rho(x) = \alpha \exp\left ( \frac{K x}{\epsilon} \right ) 
\label{density-1d}
\end{equation} 
Alternatively, the approximate steady state expression of $\rho(x)$ in Eq. (\ref{density-1d}) can also be obtained  by writing the evolution equation for mean occupancy  \cite{freypre,ignapre}, when the excluded volume effects are ignored.  Denoting $\langle n_i \rangle  = \rho_i$ as the mean occupancy at $i^{th}$ site, the evolution equation then reads as

\begin{eqnarray}
\partial_{t} \rho_i &=& \gamma \rho_{i-1} + (1-\gamma)(1 - \delta) \rho_{i-1} \\ 
&-&  \gamma \rho_{i} -  (1-\gamma)(1 - \delta) \rho_{i} - \delta(1-\gamma) \rho_{i} \nonumber
\end{eqnarray} 
 
The corresponding steady state condition: $\partial_{t} \rho_i = 0$, leads to the relation $\rho_{i} = ( 1 -\delta + \delta\gamma) \rho_{i-1}$.  Along with the boundary condition, $\rho_{1} = \alpha$, this leads to the expression of $\rho(x)$ which is identical to Eq.\ref{density-1d}. 

Including the excluded volume effect within a MF approximation \cite{freypre,ignapre}, where we factorize the two-point correlators arising out of the product of occupation numbers of neighbouring sites would to lead to evolution equation of the form,

\begin{eqnarray} 
&\partial_{t} &\rho_i = \gamma \rho_{i-1}(1-\rho_i) + (1-\gamma)(1 - \delta) \rho_{i-1}(1-\rho_i)\\
& -&  \gamma \rho_{i}(1-\rho_{i+1}) -  (1-\gamma)(1 - \delta) \rho_{i}(1-\rho_{i+1}) - \delta(1-\gamma) \rho_{i} \nonumber
\end{eqnarray} 
The interpretation of the terms of the right are as follows: The first term corresponds to a gain term on account of a particle from a normal site at position $i-1$ hopping to site $i$, the second term is due to a particle at defect site at $i-1$ hopping to site $i$, the third and fourth term are the  terms associated with particle hopping out of site i to site $i+1$ from a normal and defect site respectively, while the last term is associated with  particle leaving the lattice from a defect site.  The corresponding steady state solution in the continuum limit ( with $ x = i\epsilon$) maybe obtained \cite{freylet,ignapre}.  Ignoring terms of the order of $\epsilon^{2}$, leads to  an implicit solution of $\rho(x)$, 

\begin{equation}
 2(\rho - \alpha) - ln \left(\frac{\rho}{\alpha} \right) = \left (\frac{x}{\epsilon} \right ) \left[ \frac{\delta(1-\gamma)}{1 - \delta + \delta \gamma}\right ]
\label{den-conti}
\end{equation}

\begin{figure}[h]
\centering
\includegraphics[width=3.4in,height=2in]{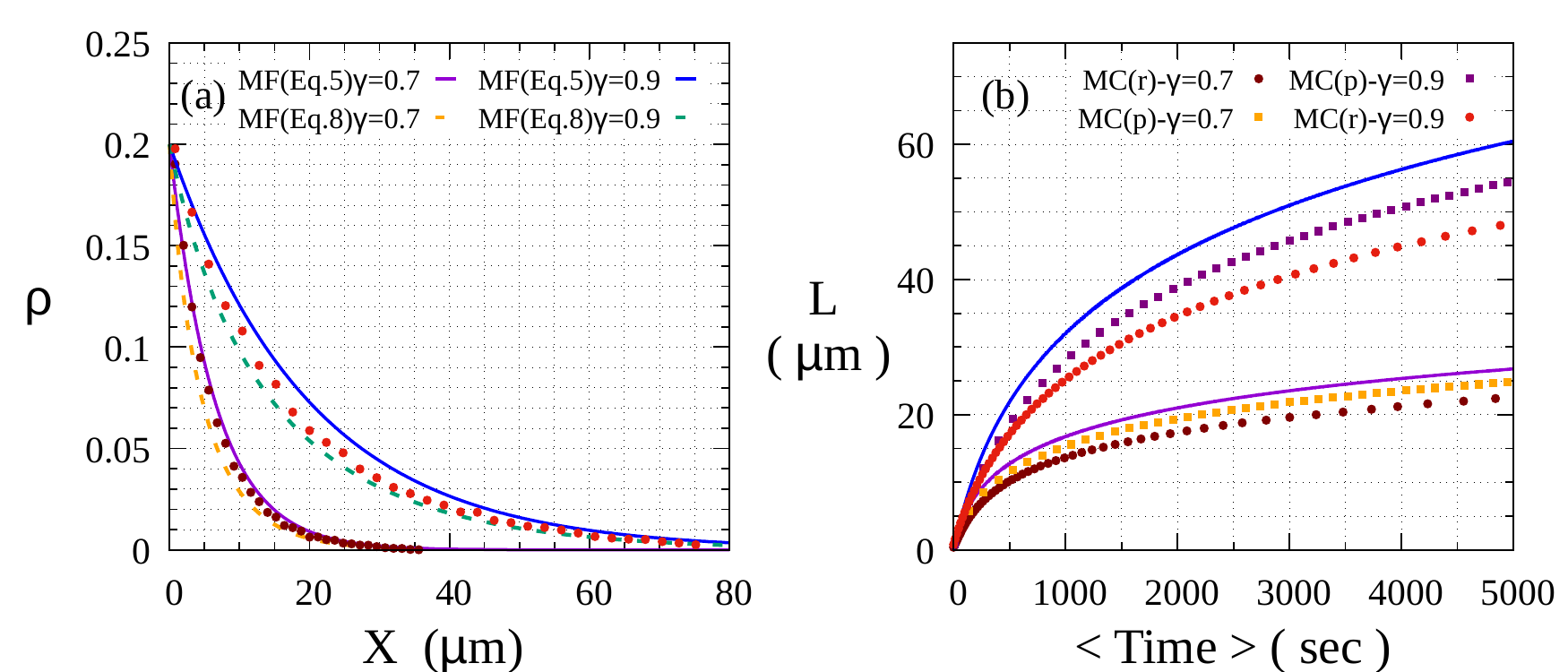}
\begin{center}
\caption{(a) Average occupancy of  particles $( \rho )$ as function of distance from the site of particle entry into the lattice $(x)$ for the 1-d model. Solid curves correspond to expression in Eq.\ref{density-1d}, while dashed curves correspond  expression of density in Eq.\ref{den-conti}. The circles correspond to points obtained by MC simulation done with random-sequential update procedure. (b) Average length of the growing lattice $ \langle L \rangle $ as a function of time $t$ : The solid curves correspond to MF expression of Eq.\ref{L}. The circles and the squares correspond to MC simulation results obtained by random-sequential update procedure and synchronous update procedure respectively. For all the plots, $\alpha =0.2 s^{-1}$, $\delta = 0.2 s^{-1}$, lattice spacing $\epsilon = 0.4 ~\mu m$, overall growth rate at tip is $0.4 ~\mu m s^{-1}$ and averaging is done for 10000 samples.}
\end{center}
 \label{fig2}
\end{figure}

In Fig.2(a), we show a comparison of spatial profile of the average occupancy along the growing 1-d lattice obtained by approximate analytical means, e.g; Eq.\ref{density-1d} and Eq.\ref{den-conti} with the profile obtained by MC simulations.

\subsection{Temporal behavior of hyphal length}
For our model we have set the rate of overall growth rate of the lattice to $1$ unit per unit time. This corresponds to a growth rate of $\epsilon$ per unit time, when the tip site is occupied by a particle. Thus the mean linear growth rate of the length of the lattice  $V \equiv d \langle L \rangle /dt$ can be expressed as, $\frac{d \langle L \rangle }{dt}  = \epsilon \rho \langle L \rangle = \epsilon \alpha \exp [ K \langle L \rangle / \epsilon ]$. With the initial condition of $L$ being $0$ at at $t= 0$, the expression for the Average length of the lattice $\langle L \rangle$ as a function of time is,
\begin{equation}
\langle L \rangle  = \frac{\epsilon}{|K|} ln ( 1 + \alpha |K| t)
\label{L}
\end{equation} 

The corresponding expression for the mean linear growth rate of the lattice length is, 
\begin{equation}
V  = \frac{ \epsilon \alpha}{1 + \alpha|K| t}
\end{equation}
In Fig.2(b), we show a comparison of the temporal profile of the average length of the growing 1-d lattice obtained by  approximate analytical means with the temporal profile obtained by MC simulations.

\subsection{ Probability distribution of Length of primary hypha}
\label{pl}
For the 1-d model for growth of the primary fungal hypha, after a fixed interval of time $t$, starting from a single lattice site, the lattice grows to $N$ lattice sites ( corresponding to a length $L =\epsilon N$). Since the process of creation of defect site is a random process, the length up to which the hypha grows is itself a random variable. Using MC simulations with synchronous update, we simulate the probability distribution of the length of the growing lattice (in terms of the total number of lattice sites N) after a fixed interval of time. 
Fig.3(a) displays the probability distribution of $N$ for different value of $t$. For this distribution the actual value of the peak of the distribution is close to approximate expression of average length obtained from Eq. \ref{L}. Fig.3(b) displays the probability distributions function of the scaled variable $z = ( N - \langle N \rangle ) / \sqrt{\langle N \rangle }$, at different times. This distribution tends to a Gaussian distribution. The width of the distribution in terms of scaled variable $z$ decreases very slowly ( $\sim 5\%$, when $t$ increase by one order of magnitude). This implies that the relative width of the distribution function of $N$, and equivalently the relative width of the probability distribution function of $L$ decreases more rapidly than $L^{-1/2}$. Thus at late times, the variation of the length of the hypha from average value is small. 

\begin{figure}[t]
\centering
\includegraphics[width=1.0\linewidth]{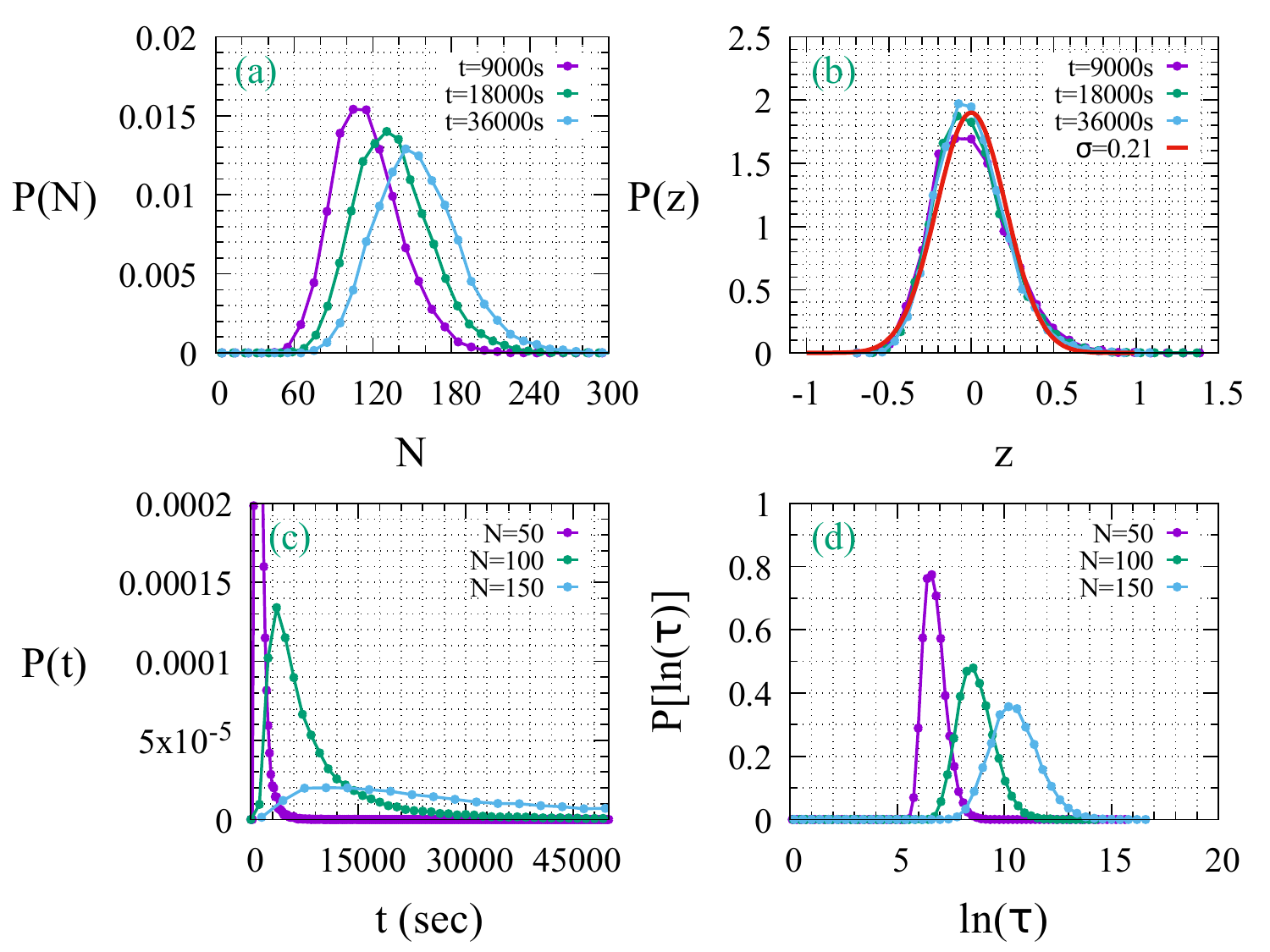}
\begin{center}
\caption{(a), Probability distribution function  of $N$ (corresponding to a length $L = \epsilon N$ of the growing 1-d lattice) after time $t$, starting from one lattice site (N =1)  at $t=0 s$. (b) Probability distribution function in terms of scaled variable $z=(N-<N>)/\sqrt{<N>}$ at different times. The solid curve corresponds to a Gaussian distribution with $\sigma = 0.21$.  (c) Probability distribution function of time $t$  required for growth of the lattice from $1$ lattice site to $N$ lattice sites. (d) Probability distribution of the $ln (\tau) $ where $\tau$ is a dimensionless parameter defined as $ \tau  = t / 1s$. For all the plots, $\alpha=0.2$, $\gamma=0.9$, $\delta=0.3$ and the probability distribution functions are obtained by averaging over 5000 samples.}
\end{center}
 \label{fig3}
\end{figure}

 \begin{figure*}[t]
    	\centering
    	\includegraphics[width=0.9\linewidth]{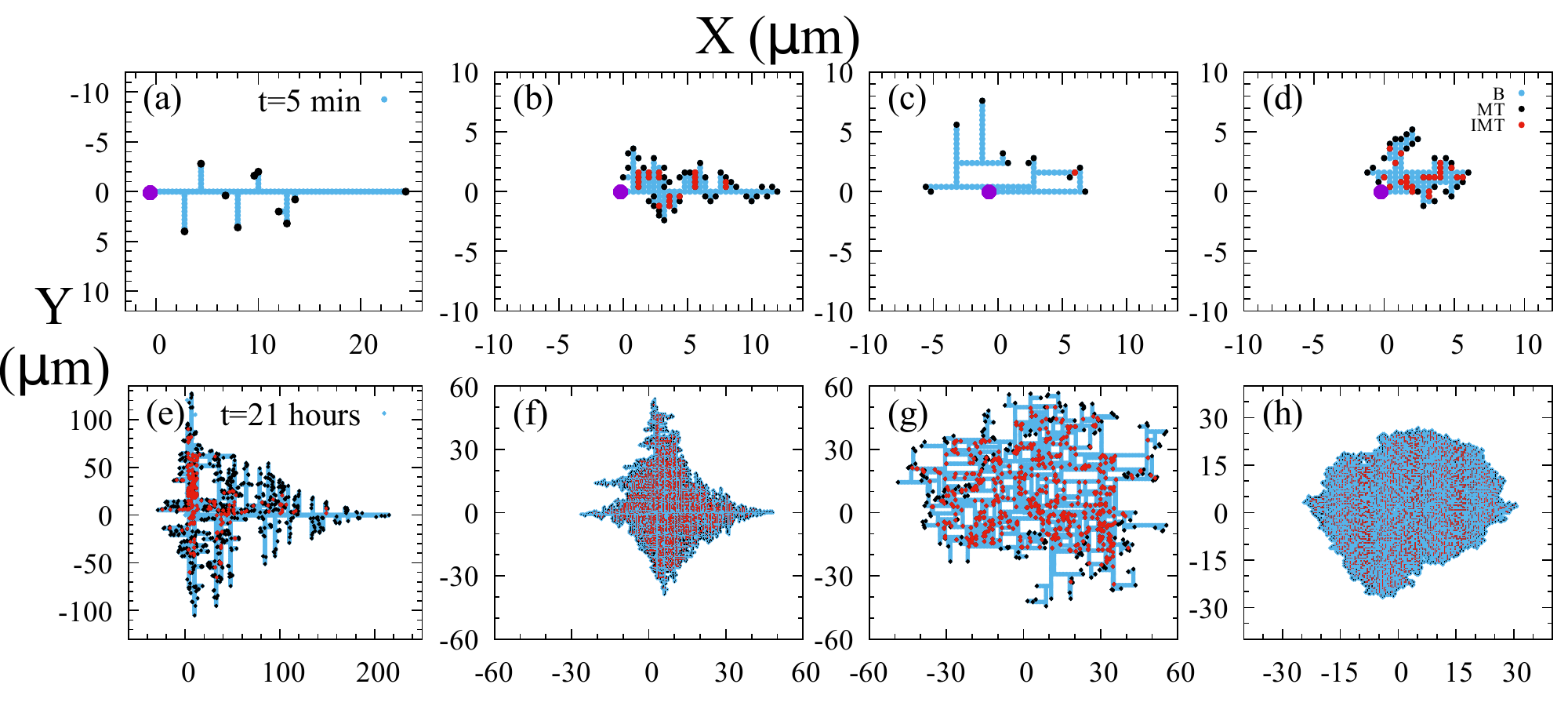}
    	\caption{The morphology of single colony mycelium (a single sample) at different times $t$, starting from an initial configuration of 2 lattice sites at $t =0$.. Area enclosing each lattice site is a square box with sides of length $0.4 \mu m$. Panels, (a) and (e) shows the morphology at different times ( $ t = 5 mins$ and $t = 21$ hours), when $\gamma=0.9,\delta=0.1$, i.e; for low branching rate and low rate of flow of nutrients to secondary branches. Panels (b) and (f) are the morphologies  at different times for $\gamma=0.2,\delta=0.1$, i.e; for high branching rate and low rate of flow of nutrients to secondary branches. Panels (c) and (g) are the morphologies  at different times for $\gamma=0.9,\delta=0.8$, i.e; for low branching rate and high rate of flow of nutrients to secondary branches. Panels (d) and (h) are the morphologies at different times for $\gamma=0.2,\delta=0.8$, i.e; for high branching rate and high rate of flow of nutrients to secondary branches. For all cases $\alpha=0.2$. The mobile tip sites (MT), immobile tip sites (IMT) and bulk sites (B) of growing colony are indicated in each panel.}	
    	\label{fig4}
    \end{figure*}

\subsection{ Probability distribution of growth time of hypha}
\label{pt}

We look at the probability distribution of time $t$ required for growth of the primary hypha to grow up to a fixed length. Fig.3(c) displays the probability distribution function of time $t$ for different values of total number of lattices sites $N$. The probability distribution function of time is generically a {\it broad distribution}, with the most probable value of $t$ significantly different from the mean value of time required for growth of the primary hypha. In Fig. 3(d) we plot the distribution of the $ln (\tau) $ where $\tau$ is a dimensionless and defined as $ \tau  = t/ 1s$. This distribution tends to a Gaussian distribution function indicating that $P(t)$ tends to a log-normal distribution function. 

\section{Features of Single Colony Mycelium}
\label{sec-2d}

We now focus our attention on the motility and morphology characteristics of a single colony mycelium. We use the 2-d model described in Subsection~\ref{2-d} for mimicking the growth processes in single colony mycelium which results from elongation and lateral branching of a single primary hypha with a single source of nutrient. The morphology characteristics, e.g; shape and size of the single colony mycelium is determined by the parameters $\gamma$ which characterizes the propensity for branching in the hypha, $\delta$ which is a parameter that characterizes the nutrient flow rate to the secondary branch and input rate of nutrients $\alpha$. Depending on the choice of these parameters, the morphology characteristics can vary significantly. 
For experiments with {\it Nuerospora Crassa}, the experimentally observed growth rates of hypha are in the range of  $20-30 \mu m /min$ \cite{gero, neurospora}. Consistent with this observation we set the overall growth rate in our model system to $0.4 \mu m/s$ and choose a lattice spacing $\epsilon = 0.4 \mu m$, which corresponds to overall growth rate of $1 s^{-1} $. This would also imply that the particles in the bulk of the lattice are hopping with rate $0.4 \mu m  s^{-1}$ corresponding to hop rate of $1 s^{-1}$ that we have set for our model. This choice of hopping rate is comparable to the typical motor velocity of kinesin-1 motors whose velocity is $\sim 0.6 \mu m s^{-1} $ \cite{gero}. Fig.4 displays some of the resultant morphologies and their evolution over time, for different set of parameters. As would be expected, when both the branching rate $ 1- \gamma$ and nutrient flow rate to the secondary branch $\delta$ is low, the lateral growth of hyphae is small compared to the longitudinal growth along the direction of growth of the primary hypha, in the initial phase of evolution of the single colony (Fig.4a) and it continues to persist after $21$ hours (Fig.4e). Fig.4b and Fig.4f shows the temporal evolution of the morphology when the branching rate is high and rate of nutrient supply to secondary branches is low, while Fig.4(c) and Fig.4(g) shows the typical morphology when the branching rate is low while the rate of supply to the secondary branches is high. When both the branching rate and nutrient flow rate to the branches is high, asymmetry of growth along the longitudinal and lateral directions is virtually absent and the single colony mycelium tends to a radially symmetric configuration about its Center of Mass at late times (Fig.4h). 

The process of branching in the hyphae has the effect of creating new tip sites (T) which in turn serve as seeds for further elongation and branching process of the mycelium. However as described in subsection \ref{2-d}, the tips can elongate or branch {\it only} if the adjacent space is not occupied by the single colony mycelium. For the case  where the branching rate and $\delta$ is high, not only would there be numerous branches which would be created in the mycelium, but there would be a tendency of the growth direction of the tips to change and consequently be obstructed by the pre-existing mycelium. Thus over the course of evolution of the morphology, many of the growing tips (T) would become immobile or jammed, i.e; these tips would not serve as either elongation or branching site. Along with  the different morphologies obtained for the different values of control parameter $\gamma$ and $\delta$, Fig.4  also shows the immobile and mobile tip sites of the mycelium after time $t$ has elapsed starting from a single hypha of unit length.  

\subsection{Motility behavior of Center of Mass}
\label{centerofmass}
In Fig.5(a) we display the temporal behavior of the Center of Mass of the growing lattice network (averaged over different samples). Increasing the branching rate $1- \gamma$ and/or $\delta$ has the effect of impeding the motility of the center of mass of the single colony mycelium. As expected for early times, for low $\delta$ and low branching rates, $\sqrt{\langle R_{cm}^2\rangle }$ shows a linear dependence on the $\tau$. From the log-log plot in Fig.5(b), we can infer that $\sqrt{\langle R_{cm}^2\rangle} \sim  t^{q} $ at later times. For all the cases we find that the exponent $q  <1/2$ indicating sub-diffusive behavior of the Center of mass for the regime of time window that we have probed. Fig.5(c) shows the dependence of $\langle R_{cm}^2\rangle$ on $N$.  From the log-log plot displayed in  Fig.5(d) for different set of input parameters we see that for sufficiently large size N, $\sqrt{\langle R_{cm}^2\rangle} \sim  N^{\beta}$.

\begin{figure}[h]
 \centering
   \includegraphics[width=1.0\linewidth]{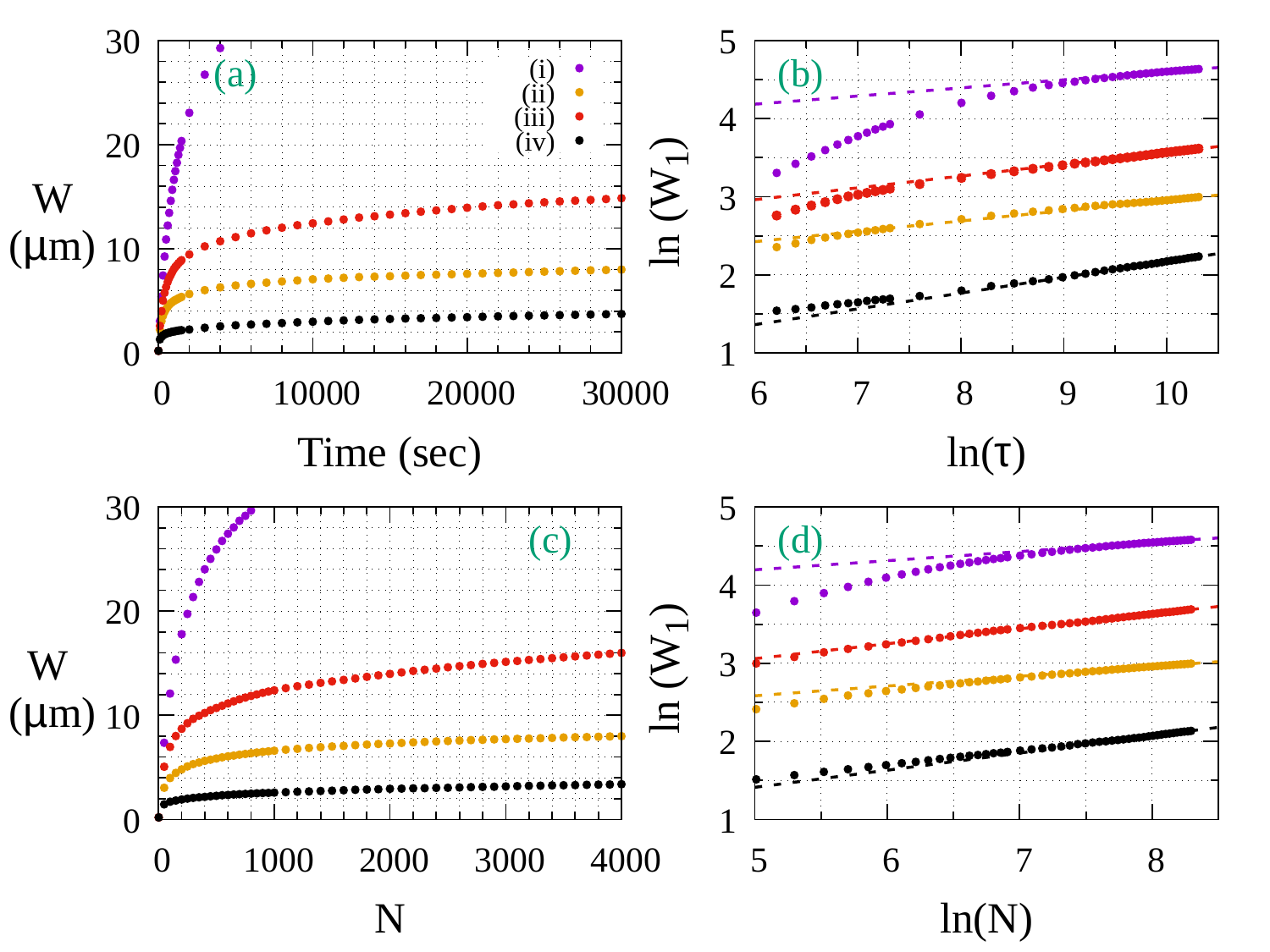}
    \caption{Behavior of Center of Mass $ W \equiv \sqrt{\langle R_{cm} ^{2} \rangle}$ for four  different parameter values: (i) $\gamma=0.9, \delta=0.1$, (ii) $\gamma=0.2, \delta=0.1$, (iii) $\gamma=0.9, \delta=0.8$ and (iv) $\gamma=0.2, \delta=0.8$. For all cases $\alpha = 0.2$ and $\epsilon = 0.4 \mu m$. These set of parameter values are the same as in Fig.\ref {fig4} (a), (b),(c) and (d) respectively.  Panel (a) displays the plot of $W$ vs $t$. (b) Plot of $log (W_1)$ vs $log (\tau)$, where $W_1 = W / \epsilon$ is the dimensionless Center of Mass, and $\tau = t/ 1sec$, is the dimensionless time. The least square fitted straight line to the data points at  later times corresponds to a slope of  (i) $q = 0.10$,  (ii) $q = 0.13$, (iii) $q = 0.15$ and (iv) $q = 0.20$ where $ W_1 \sim \tau^{q}$. (c) Plot of $W$ vs $N$. (d) Plot of $log (W_1)$ vs $log (N)$ : The least square fitted straight line to the data points at large N corresponds to a slope of  (i) $\beta = 0.12$,  (ii) $\beta= 0.12$, (iii) $\beta = 0.19$ and (iv) $\beta = 0.22$ where $ W_1 \sim N^{\beta}$. Averaging is done over $100$ samples.}
    \label{fig5}
  \end{figure} 
\begin{figure}[h]
 \centering
   \includegraphics[width=1.0\linewidth]{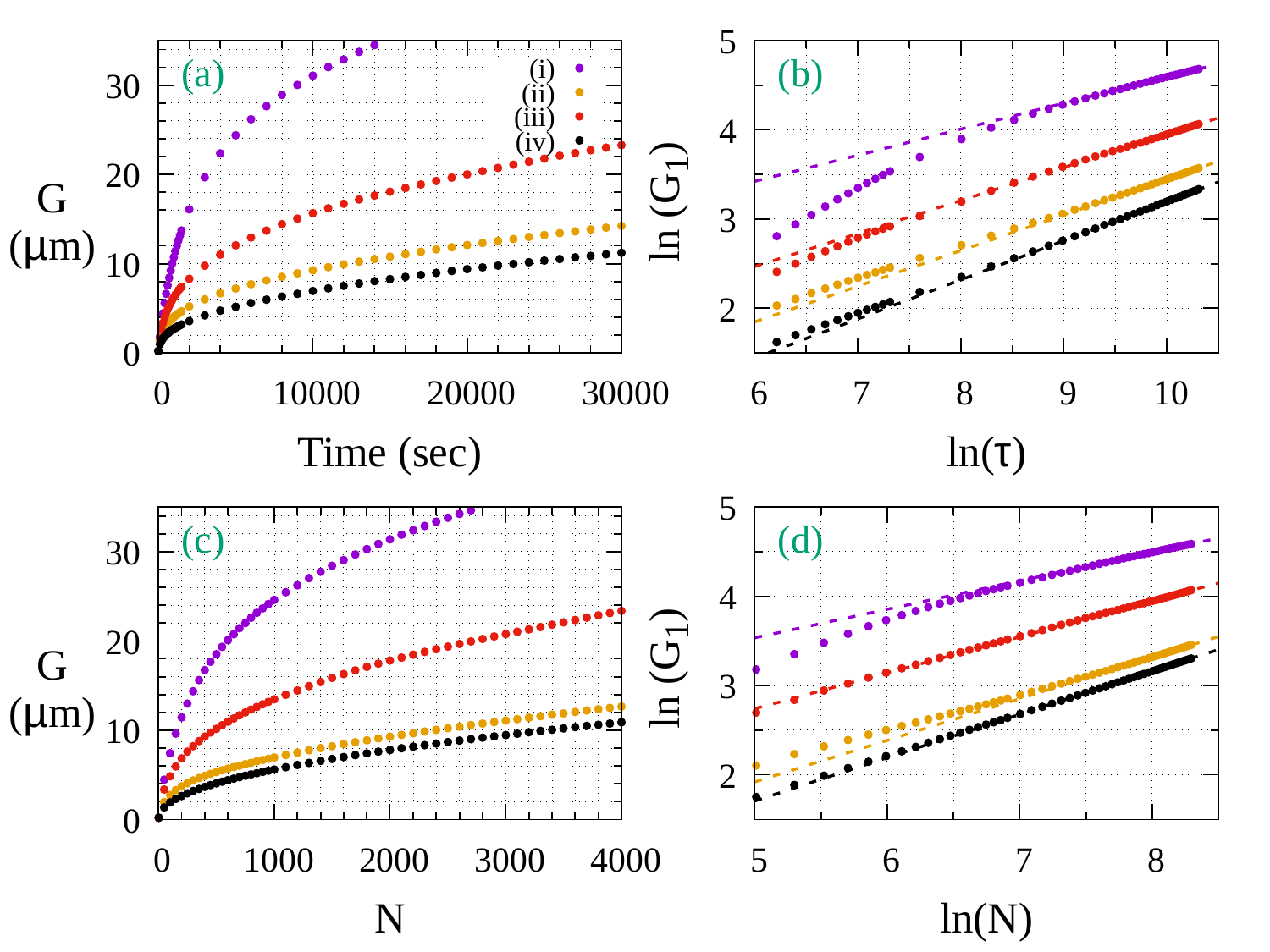}
    \caption{ Behavior of Radius of Gyration $G =\sqrt{ \langle R_{G}^{2} \rangle}$ for four different parameter values (same as Fig.5). Panel (a) displays the plot of $G$ vs $t$. (b) Plot of $log (G_1)$ vs $log (\tau)$, where $G_1 = G / \epsilon$ is the dimensionless Radius of Gyration, and $\tau = t/ 1sec$, is the dimensionless time. The least square fitted straight line to the data points at  later times corresponds to a slope of  (i) $\phi = 0.29$,  (ii) $\phi = 0.40$, (iii) $\phi = 0.37$ and (iv) $\phi = 0.44$ where $ G_1 \sim \tau^{\phi}$. (c) Plot of $G$ vs $N$. (d) Plot of $log (G_1)$ vs $log (N)$ : The least square fitted straight line to the data points at large N corresponds to a slope of  (i) $\nu = 0.32$,  (ii) $\nu= 0.47$, (iii) $\nu = 0.40$ and (iv) $\nu = 0.48$ where $ G_1 \sim N^{\nu}$. Averaging is done over $100$ samples.}
    \label{fig:2a}
  \end{figure}

\subsection {Size of the Colony} 
\label{size}

Radius of Gyration $R_g$ is one measure which can be used to characterize the typical size of the single colony mycelium. In order to understand the growth characteristics of the size of the single colony we look at the temporal behavior of  $\sqrt{\langle R_{g}^2\rangle}$. Fig.6(a) displays the temporal evolution of $\langle R_{g}^2\rangle$. From the log-log plot in Fig.6(b) we can infer that at later times, $\langle R_{g}^2\rangle \sim  t^{\phi}$ with $\phi < 1/2$. Fig. 6(c) displays the variation of $\langle R_{g}^2\rangle$ with $N$ for different set of input parameters.  Assuming a power law dependence of $\sqrt{\langle R_{G}^2\rangle}$ on $N$ of the form $\langle R_{g}^2\rangle \sim  N^{2 \nu}$, from the log-log plot in Fig.6(d) we extract the exponent $\nu$ from the slope of the least square fitted straight line. For the range of $N$ considered for our simulations, the exponent $\nu$ varies with the chosen inputs parameters $\delta$ and $\gamma$. Typically the effect of lowering the branching rate has the effect of increasing $\langle R_{G}^2\rangle$ for a given $N$. When nutrient supply rate $\delta$ is made high while branching rate is lowered, the  resultant colony becomes more {\it porous}, with less dense filling of space. This attribute of the colony is demonstrated by the morphology of the colony shown in Fig. 4(g). This maybe understood as follows: When the branching rate is low, very few branches are created. However the the secondary branches are the ones which grows faster. Subsequent branching of these secondary branches eventually results also to formation of closed loop, which arrest further growth of the branches locally due to obstruction. The size of this loop determines the porousness of the colony. Since the size of these closed loops increases when branching rate is decreased and $\delta$ increased, the resultant overall morphology becomes more porous. When both the branching rate is high and $\delta$ is high, it would be expected that the porousness would be low and the growth of the colony would be a space filling process. Such a scenario is observed For Case (iv), for which  $\gamma=0.2, \delta=0.8$. The typical configuration of the colony for this particular choice of parameters is illustrated in Fig. 4(h). For this case, the corresponding value of the exponent $\nu \simeq 0.5$ in the large $N$ limit. This value of the exponent $\nu$ is same as that of 2-d Eden model for growth in the large $N$ limit \cite{eden1,dhar1}. Further for large $N$ limit, $\langle R_{G}^2\rangle \rightarrow N/2\pi$, which is again similar to the results obtained for the 2-d Eden Model \cite{eden1}. In fact any 2-d growth process which incorporates excluded volume effect and allows for maximum occupancy of $1$ per lattice site would necessarily have to satisfy the condition, $\langle R_{g}^2\rangle \geq N/2\pi$. This condition arises purely due to geometric constraint. The equality holds for the case of complete space filling on a circle. Except for Case (iv) for which  $\langle R_{g}^2\rangle$ approaches the limiting value of  $N/2\pi$, for the other three cases,  $\langle R_{g}^2\rangle >  N/2\pi$ (Fig.6c) up to $N =4000$. Further,the value of exponent $\nu < 0.5$ for the other three cases which implies that the value of the exponent $\nu$ characterizing the power law behavior of the Radius of Gyration ( up to $N=4000$) is different from the exponent $\nu$ which characterize the truly 'large' $N$  limiting behavior of $\langle R_{g}^2\rangle$.     

\subsection {Shape Characteristics of the Colony}
\label{shape}

  \begin{figure}[h]
 \centering
 \includegraphics[width=2.5in, height=2in]{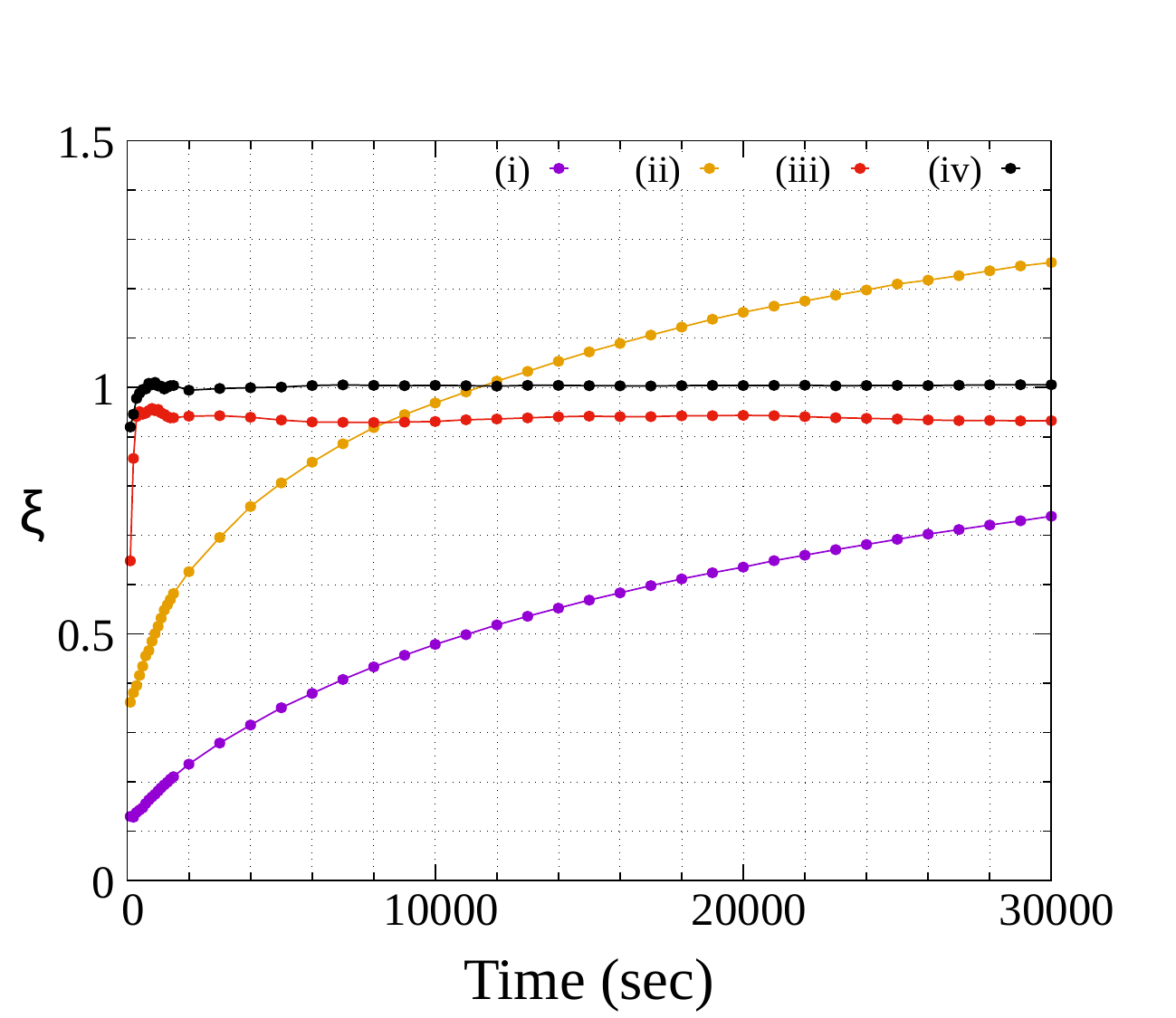}
    \caption{ Temporal evolution of Aspect ratio  $\xi = \sqrt{ \langle Y_{g}^{2} \rangle / \langle X_{g}^{2} \rangle}$: Plot of $\xi$ vs $t$ for four different parameter values (same as Fig.5).  Averaging is done over $100$ samples.}
    \label{fig:2a}
  \end{figure}

 In order to characterize the asymmetry of the morphology of the single colony we define aspect ratio $\xi = \sqrt{ \langle Y_{g}^{2} \rangle / \langle X_{g}^{2} \rangle} $.  $\xi$ is a measure of the relative growth in the lateral direction viz-a-viz the extension along the original direction of growth of the primary hypha. Fig. 7 shows  the temporal behavior of the aspect ratio. For very late times, the aspect ratio is expected to converge to 1. Indeed, when the branching rate is low and nutrient supply rate to the secondary branches is high, as in  Case (iv), the aspect ratio of the colony tends to 1 very rapidly which corresponds to a typical morphology of the colony shown in Fig. 4(h). However when the branching rate is low and $\delta$ is also low, for early times, the lateral extension of the colony ( perpendicular to initial growth direction of the original hypha) is much less compared to growth along the original direction along x. Thus even after more than 8 hours of growth, the aspect ratio is much lower than 1, approaching 'late' time limit of $\xi =1$  slowly.  Interestingly for a range of intermediate times ( up to $30000$ secs ) considered in the simulation, when the branching rate is very high, while the nutrient supply rate to the branches is low, as in Case (ii),  the aspect ratio $\xi$ even exceeds $1$ (Fig. 7). This maybe qualitatively be understood as follows: When the branching rate is high, whilst $\delta$ is low, from the primary hypha, many lateral branches(along y axis) would be generated, and the typical distances between these lateral branches would be relatively small. Over the course of evolution of the colony, these lateral branches would be the sites for secondary branches along $x$ axis- the direction of growth of the original primary hypha. However the growth of these secondary branches would be arrested by the frequent lateral branches arising due to high rate of branching. This is turn would lead to a situation where the overall growth along the x direction(except the primary branch) would be slowed down compared to the unhindered growth of the secondary branches along the lateral direction along y axis. Such growth process results in a typical morphology of the colony shown in Fig 4(f) at intermediate times. However even for this case, at later time $( \sim 130 $ hours), $\xi$ converges to 1.

\subsection{Growth sites of the colony}
\label{tips}

For the 2-d model for the growing colony that we have considered, growth occurs only at the tip sites. While the elongation process shifts the position of the tip site, the branching process results in creation of new tips. Over the course of evolution of the morphology of the colony, many of the tip sites cease to be sites of growth and become immobile as they are obstructed by the part of the pre-existing colony. Fig.8 displays the temporal behavior of $Q$ which is the ratio of the number of mobile tips $N_m$, with total number of tips $N_T$ in a single colony. $Q$ decreases with the passage of time as the colony size increases. Expectedly, the rate of decrease of $Q$ is faster for the case when the branching rate is high and the nutrient supply rate to secondary branches is also high (Case iv) in comparison to the case when both the branching rate and $\delta$ are low (Case i). In fact, at 'late' times, and 'large' N limit, the mobile tips would reside typically on the outer perimeter of the colony while the immobile tips would be part of the bulk (Fig.4h). As the size of the colony grows, the ratio of the number of sites on the outer perimeter with the total number of sites, decreases and consequently we would expect the decrease in $Q$ as we observe in the simulations. In order to quantify the typical scaling behavior exhibited by these mobile tips in the 'large' N limit, we specifically focus on the case where branching rate is high and for which $\delta$ is low, since for this case convergence to 'large' N limit behavior is relatively faster. We define $S$ as the ratio of the number of mobile tips $N_m$ with the total number of lattice sites of the colony $N$. In Fig. 9(a) we show the variation of $S$ with Radius of Gyration $R_{g}(N)$. From the corresponding log-log plot in Fig.9(b), we can infer $S \sim R_{g}^{-1}$. This particular scaling behavior maybe understood as follows: For large N limit, the mobile tips reside on the outer perimeter of the colony. Further if the perimeter is an euclidean surface, then the ratio of the perimeter length with Total area should scale inversely with $R_g$ and consequently $S$ itself scales inversely with the Radius of gyration. Since for this case, from our simulations, we know that $R_g \sim N^{1/2}$ (Fig.6, Case iv), it would be expected that $S \sim N^{-1/2}$. Fig.9(c) displays the variation of $S$ with $N$. The corresponding log-log plot displayed in Fig.9(d) indeed confirms this scaling behavior.   

\begin{figure}[h]
 \centering
    \includegraphics[width=2.5in, height=2in]{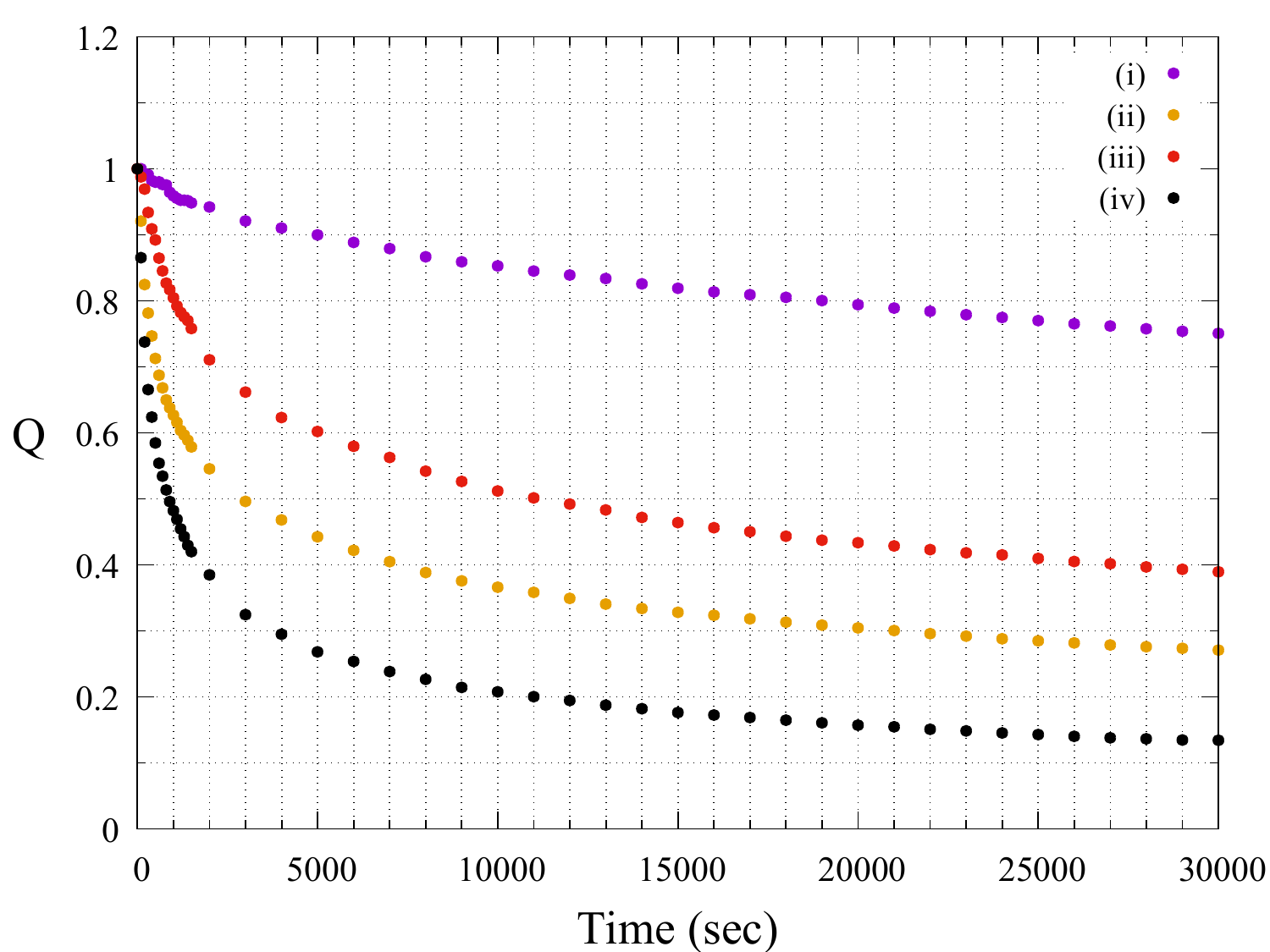}
    \caption{Temporal behavior of the mobile tip ratio $Q \equiv N_{m} / N_T$, where $N_m$ is the number of mobile tips and $N_T$ is the total number of tips.  Variation of $Q$ for four different parameter vales ( same as Fig.5) is shown. Averaging is done over $100$ samples.}
    \label{fig:2a}
  \end{figure}

\begin{figure}[h]
 \centering
    \includegraphics[width=3in, height=3.2in]{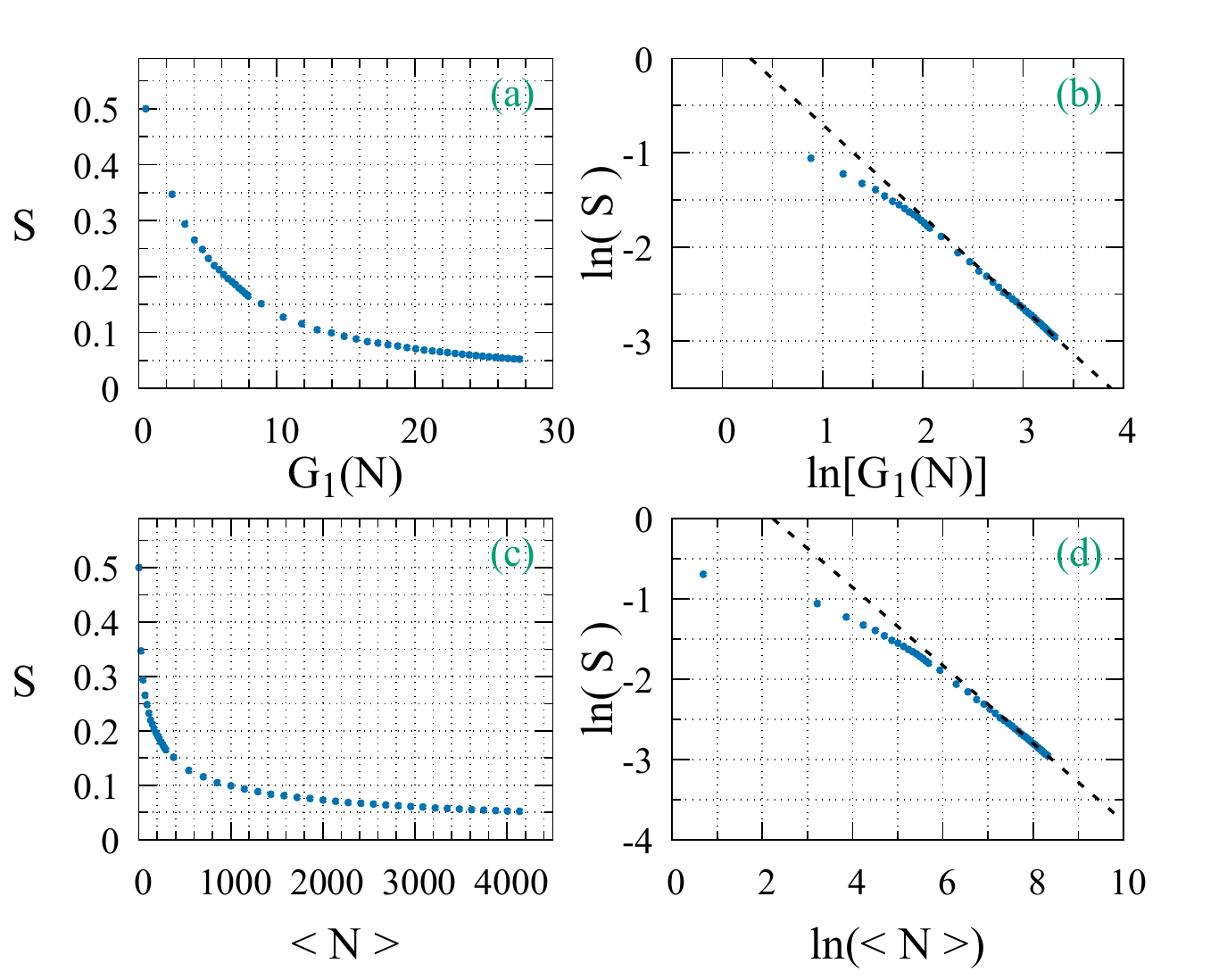}
    \caption{Scaling behavior of $S \equiv \langle N_m \rangle / \langle N \rangle $: (a) Variation of S with Scaled Radius of Gyration $G_1$, (b) Plot of $log (S)$ vs $log (G_1)$:  The least square fitted straight line to the data points at large $N$ corresponds to a slope of  $\theta = - 1.03$, where $ S \sim G_{1}^{\theta}$. (c) Variation of $S$ with $\langle N \rangle $. (d) Plot of $log (S)$ vs $log(\langle N \rangle )$:  The least square fitted straight line to the data points at large $N$ corresponds to a slope of  $\eta = - 0.49 $, where $ S \sim N^{\eta}$. Here $\gamma=0.2, \delta=0.8$ and $\alpha = 0.2$. Averaging is done over $100$ samples.}
    \label{fig:2a}
  \end{figure}

\section{Comparison with Experiments}
\label{sec-expt}

The only control parameters of the 2-d minimal model that we have presented are the entry rate $\alpha$, the branching rate $1 - \gamma$, and nutrient supply rate to secondary branches $\delta$. In order to make connection with the morphological features observed in experiments, we check whether by calibrating these three input parameters, the model is able to reproduce quantitative measures associated with the growth characteristics of fungal hyphae. Consistent with experiments with fungal hyphae of {\it Nuerospora Crassa}, we choose overall hyphal extension rate $0.4 \mu m/s$ \cite{neurospora, gero} and set the lattice spacing $\epsilon = 0.4 \mu m$. This would correspond to an overall growth rate of $1$ lattice unit per second. For experiments with {\it Nuerospora Crassa}, the experimentally observed growth rates of hypha are in the range of  $20-30 \mu m$ /min \cite{gero, neurospora}. We choose a particle hopping rate of $1s^{-1}$. This would also imply that the particles in the bulk of the lattice are hopping with rate $0.4 \mu m s^{-1}$. This choice of hopping rate compares reasonably with the typical motor velocity of kinesin-1 motors \cite{gero}. 

\begin{figure}[h]
 \centering
    \includegraphics[width=1.0\linewidth]{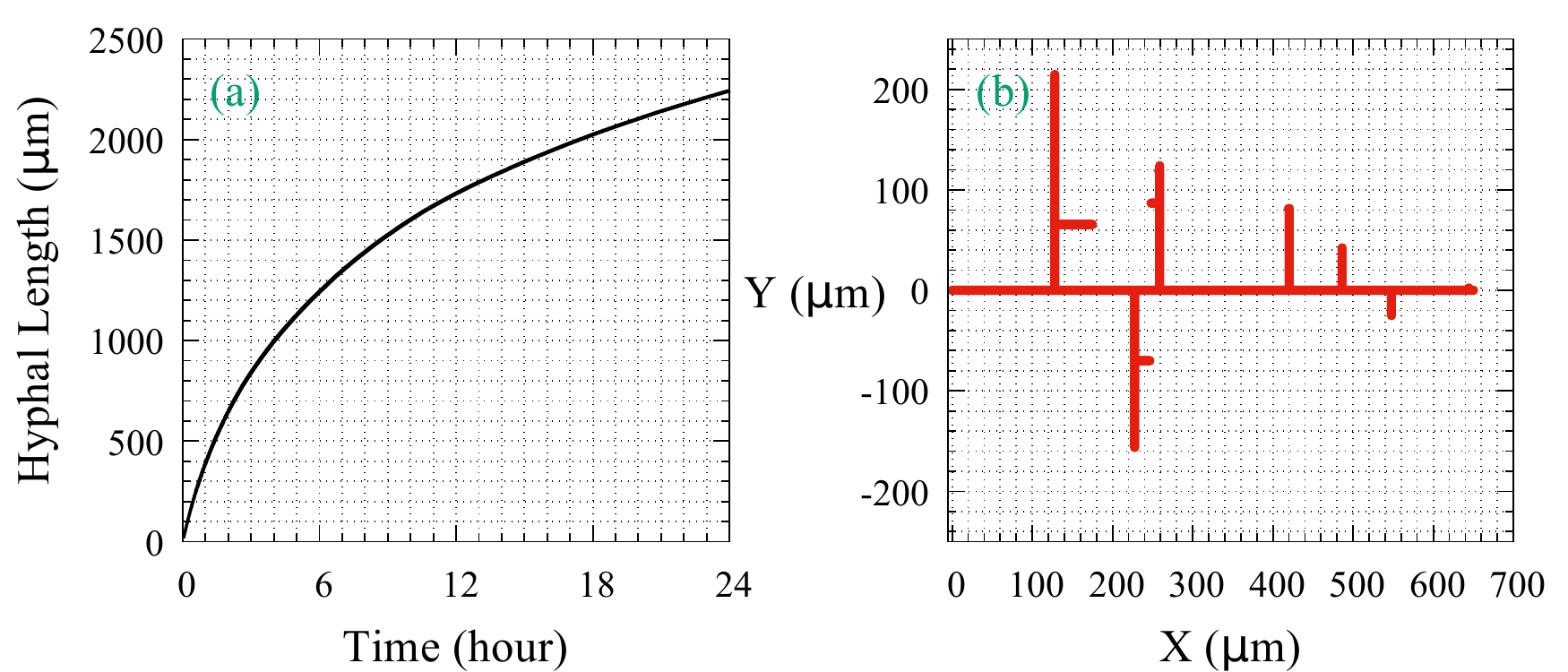}
    \caption{(a) Average length of lattice $ \langle L \rangle $ as a function of time $t$ : The solid curves correspond to MC simulation results for 1-d model. Here, $\alpha =0.4 s^{-1}$, $\delta = 0.2 s^{-1}$, $\gamma = 0.998 s^{1}$, lattice spacing $\epsilon = 0.4 \mu m$ and velocity  at the growing tip is $0.4 \mu m s^{-1}$. For these choice of parameters, the hyphal growth characteristics observed experimentally for fungal hypha of  {\it Rhizopus Oligoporous} ( Ref.\cite{expt, plos}), matches reasonably well with the MC simulation results. (b) The 2-d morphology after $5$ hours obtained for $\alpha =0.4 s^{-1}$, $\delta = 0.2 s^{-1}$, $\gamma = 0.995$, $\epsilon = 0.4 \mu m$ and velocity at the tip is $0.4 \mu m s^{-1}$. This morphology is similar to the morphology of the single colony hypha obtained in experiments with {\it Nuerospora Crassa} ( Ref.\cite{neurospora}).}
    \label{fig:2a}
  \end{figure}

In Fig.10(a) we plot the temporal behavior of the average length of the primary hypha, using MC simulations of the 1d model discussed in subsection \ref{1-d} for a particular choice of input parameters. This profile quantitatively reproduces reasonably well the temporal profile obtained experimentally for the hyphal growth  of {\it Rhizopus Oligoporous} \cite{expt,plos}.

In Fig.10(b) we show the morphology of a single sample obtained by MC simulations of the 2-d model discussed in subsection \ref{2-d} for a particular choice of parameters $\alpha$, $\gamma$ and $\delta$ after $5$ hours of growth. This simulated morphology compares well with the morphology observed in experiments with {\it Neurospora Crassa} after 5 hours of growth \cite{neurospora}.

\section{Conclusions and Discussion}
\label{sec-conclusion}
In this article, we have discussed a minimal driven lattice gas model which generates the morphological characteristics associated with single colony mycelium arising from the growth and branching process of fungal hyphae, which is fed by a single source of nutrients. While the 1-d model describes the growth characteristics of the primary hypha, the 2-d model provides a description of the entire single colony mycelium that is generated by the elongation and branching process of the fungal hyphae. 

The 1-d model predicts a spatial profile of particles which is exponential along the direction of growing primary hypha, and a length of the primary hypha which grows logarithmically with time. Our MC simulation results shows that the sample to sample relative fluctuation of the average length of the primary hypha at late times is small. However we find that the probability distribution of the time required to grow to a specified length is broad and it tends to a log-normal distribution at late times. Although we have not been able to explicitly derive an explicit analytical from of the distribution function of the growth time, we have been able to show that the analytical form of the survival probability of  particle to reach distance $L$ along the primary lattice without getting lost from the primary lattice is a log-normal distribution. As an aside, it maybe noted that this kind of log-normal distribution with its implication of huge sample to sample variation of the measured growth time is seen also in context of  probability distribution function of tangent- tangent correlation function of a random heteropolymer \cite{madan-jstat} and spin-spin correlation function for random Ising spin chain \cite{derrida-lognormal}.

Using MC simulations, we generate the different types of 2-d morphologies of single colony mycelium. We find a wide variability of motility, size and shape characteristics of the 2-d morphology depending on the input parameters, e.g;  branching rate $1 - \gamma$, nutrient flow rate to secondary branches $\delta$ and input rate of nutrients $\alpha$. The Center of Mass $R_{cm}$ and the Radius of Gyration $R_g$, of the simulated colony exhibit a sub-diffusive behavior, at later times. Our analysis also reveals a power law dependence of $R_{g}$ on $N$ of the form $R_{g}\sim  N^{\nu}$. 
When both the branching rate is high and $\delta$ is high, for sufficiently high $N$, the behavior of $R_g$ as a function of $N$ is characterized by exponent which seems to converge to $\nu \sim 0.5$, which is similar to the value of the exponent that characterizes the growth process of an Eden cluster in the large $N$ limit \cite{eden1,dhar1}. A more detailed investigation is needed to confirm whether growth process of our 2-d model, in the limit of high branching rate and high $\delta$ belongs to the same universality class as that of the Eden model \cite{eden,eden1,dhar1}. More generally, it would be interesting to study the similarities and the differences of the asymptotic characteristics of the growing colony at large $N$ and late time limit for this model with other lattice animal models  of growth \cite{eden1, dhar1,latticeanimal1} and polymerization process of branched polymers \cite{polymer,lubensky1,lubensky2, rosa1, rosa2}.

We find that by suitable calibration of the parameters of the model, we are able to quantitatively reproduce the observed experimental growth characteristics of the primary hypha of {\it Rhizopus Oligoporous} \cite{plos,expt}, and we are also able to replicate the morphology characteristics of the single colony mycelium of {\it Neurospora Crassa} observed in experiments \cite{neurospora}. 

While in this article, we have restricted ourselves to analyzing the morphology characteristics of the colony for which the tip site stops growing and branching when it encounters obstruction due to the presence of the already grown lattice, it would interesting to study the morphology characteristics of the colony for the situation for which parts of the growing network is allowed to overlap over each other to form a layered mesh.

\vskip 0.5cm
\noindent
{\em Acknowledgements.} Financial support is acknowledged by SM for SERB project No. EMR/2017/001335. SM would like to thank D.Dhar (IISER, Pune) and Ignacio Pagonabarraga (CECAM, Switzerland) for useful discussions and suggestions.

\vskip 0.5cm
\noindent
B.S carried out simulations and analyzed data. S.K carried out simulations for the 1-d model. S.M designed the study, analyzed data and wrote the manuscript.

\bibliographystyle{prsty}

\end{document}